# A SCHEME FOR MAXIMAL RESOURCE UTILIZATION IN PEER-TO-PEER LIVE STREAMING

Bahaa Aldeen Alghazawy and Satoshi Fujita

Department of Information Engineering, Hiroshima University, Hiroshima, Japan

## ABSTRACT

*Peer-to-Peer streaming technology has become one of the major Internet applications as it offers the opportunity of broadcasting high quality video content to a large number of peers with low costs. It is widely accepted that with the efficient utilization of peers and server's upload capacities, peers can enjoy watching a high bit rate video with minimal end-to-end delay. In this paper, we present a practical scheduling algorithm that works in the challenging condition where no spare capacity is available, i.e., it maximally utilizes the resources and broadcasts the maximum streaming rate. Each peer contacts with only a small number of neighbours in the overlay network and autonomously subscribes to sub-streams according to a budget-model in such a way that the number of peers forwarding exactly one sub-stream will be maximized. The hop-count delay is also taken into account to construct a short depth trees. Finally, we show through simulation that peers dynamically converge to an efficient overlay structure with a short hop-count delay. Moreover, the proposed scheme gives nice features in the homogeneous case and overcomes SplitStream in all simulated scenarios.*

## 1. INTRODUCTION

With the widespread of broadband accesses to the Internet, video over IP has attracted more and more users in recent years. For example, it is forecasted that internet video streaming and downloads will grow up to more than 80% of the global internet consumer traffic by 2019 [1]. Those streaming services can be efficiently supported by IP multicast, but unfortunately, this service is not widely deployed until now for many reasons. Some of them are: the current multicast model is very costly in term of installation and management, the lack of supporting some functionalities like group management, security and address allocation, and the lack of a good pricing model. As an alternative technique, video streaming over Peer-to-Peer (P2P) network has attracted considerable attention in recent years [2,3,4,5,6,7,8]. Moreover, content delivery networks are assisted by P2P networks to reduce the economic cost of broadcasting a video to a high number of viewers [9,10]. The basic idea of P2P video streaming is that every peer can download and simultaneously upload the video content to other peers. That means, peers contribute their resources to realize a scalable service.

Different P2P streaming systems use different overlay structures and different data dissemination protocols. In mesh-based systems [2], [3], each peer establishes neighborhood relationships to a set of random peers. Neighborhood relations may change depending on the upload capacity and the content availability of peers. Neighbors periodically exchange the content availability to "pull" missing content from each other. Such a dynamic construction of random overlay is robust against the dynamic behaviour of peers (churns), but it does not guarantee the quality of content distribution such as the delay, jitter, and the transmission overhead.

In tree-based systems [4,5,6], on the other hand, peers are organized in a tree-structured overlay and the streaming content, which is "pushed" by the media server located at the root of the tree, is delivered to the downstream peers by repeating store-and-relay operation. Tree-based systems are simple and efficient but are not robust against peer churns. That is because the failure of a peer prevents all its descendants from receiving the video till the tree is fixed. In addition, it could not fully utilize peers' resources since it does not use the upload bandwidth of leaf peers in the overlay. Such drawbacks of tree-based systems can be overcome by adopting multiple trees [7], [8]. In multiple-tree systems, peers are organized into multiple spanning trees, and the video stream is divided into multiple sub-streams; each sub-stream is delivered through a single tree. Such systems try to increase the number of peers who are a parent in only one tree and a leaf in the remaining trees. Thus, more peers will contribute their resources which significantly improve the resource utilization. Moreover, the overlay is more fault tolerant as the failure of a peer affects its descendants for only part (sub-stream) of the video.

A key issue to realize an efficient content distribution in the P2P environment is how to maximize the utilization of resources contributed by the participants. Note that by the efficient maximal utilization of resources, we can broadcast a maximal streaming rate with a short end-to-end delay. To maximally utilize the resources, we need to guarantee that all peers are engaged in the content distribution process. Then, an efficient peering and content scheduling strategy is needed which is expected to overcome both of the upload capacity bottleneck and the content bottleneck in the underlying P2P systems. Finding such a strategy becomes more difficult when the capacity of the P2P system is barely enough to broadcast the streaming content to all peers, i.e., the spare capacity to adopt future peers is very low. Although SplitStream [7] highly utilizes the upload capacity of each peer by organizing a multiple-tree overlay, it results in an inefficient overlay of degenerate trees in case of barely enough resources. That is explained in related work section and verified in our simulation.

In this paper, we adopt the multiple trees approach and propose a scheduling scheme that works in the challenging condition where no spare capacity is available. The scheme attains the maximal resource utilization while maintaining an efficient overlay of multiple short-delay trees in a distributive manner. To overcome the drawback of SplitStream, a budget-model is used in the scheduling scheme such that each peer has a budget relative to its upload capacity and corresponds to the maximum number of children that peer can have. In the scheme, the role of uploading a sub-stream is transferred to another peer by exchanging money among peers, provided that the balance of each peer is not below zero. A newly joining peer try to instantly spend its budget to subscribe to sub-streams in such a way that guarantees a high number of peers forwarding exactly one sub-stream and that the trees have a short hop-count delay. The proposed scheme is also able to broadcast the maximal streaming rate as it is able to attain the maximal resource utilization.

The performance of the proposed scheme is evaluated by simulation. The simulation result indicates that under the proposed scheme, the overlay network certainly converges to an efficient structure with a short hop-count delay. Moreover, it indicates that the proposed scheme gives nice features in the homogeneous case and overcomes SplitStream in all simulated scenarios.

The remainder of this paper is organized as follows. Section 2 overviews related works. Section 3 describes the proposed scheme. Section 4 presents simulation results. Finally, Section 5 concludes the paper.

## 2. RELATED WORK

Recently, several researchers have analysed the maximum resource utilization of peers in P2P systems with a goal to stream the maximum bit rate. In the following, we assume that the system contains a media server in addition to $n$ participating peers. In [11], an upper bound on the maximum streaming rate is derived for a fully connected network in which each peer is adjacent with any other peer. Let $u_s$ denote the upload capacity of the media server and $u_i$ denote the upload capacity of the $i^{th}$ peer. Then, an upper bound $r^{max}$ on the maximum streaming rate is given as

$$r^{max} = min\left\{u_s + \frac{u_s + U(P)}{n}\right\} \quad (1)$$

where $U(P) = \sum_i u_i$. The above formula indicates that $r^{max}$ does not exceed the upload capacity of the media server, and in addition, it does not exceed the average upload capacity of peers and the media server. Different schemes have been proposed to achieve the maximum streaming rate defined in Equation (1) in the fully connected network. In [11], the video stream is divided into uneven sub-streams depending on the upload capacity of peers. The server feeds a peer who has an upload capacity $u_i$ with a sub-stream of bitrate $u_i/n$. Each sub-stream fed by the server will be propagated to all peers in the system by repeating store-and-relay operation. However, as the number of peers increases, the bit rate of each sub-stream becomes quite low. Hence, especially for the peers with low upload capacity, it incurs an excessive transmission overhead due to a large fraction of packet headers leading to waste some resources.

In Queue based scheme [12], the stream is divided into several chunks of few kilobytes to avoid a possible transmission overhead. Those chunks are pulled / pushed from the media server to the peers, cached at forwarding queues of the receivers, and relayed from the receivers to their $n-1$ neighbors. According to the upload capacity of each peer which is "inferred" from the occupancy of its forwarding queue, the peer pulls more chunks from the server to be forwarded. Although this scheme is designed to avoid excessive transmission overhead and the bandwidth calculation, it does not avoid the overload of peers which would cause a long delivery delay. In fact, if the chunk size is $\delta$, each peer needs to forward the data of size $\delta \times (n-1)$, which easily exceeds $u_i$ as $n$ becomes large. Anyway, although both previous schemes optimally utilize the resources, it is definitely difficult to deploy the fully connected network adopted by them in the practical use. Authors in [13] studied the optimal streaming rate over general overlays with peer degree bounds (number of active connections) by using central solutions. Network coding and video coding schemes are also used in this regard. In [14], authors studied a network-coding based distributed solution to maximize the streaming rate for arbitrarily overlays and under peer degree bounds. In [15], the scalable video coding SVC is used to maximally utilize peers' resources. In this paper, we try to provide a distributed solution to achieve the maximal streaming rate where peers use only store-and-relay operations and without any coding scheme.

SplitStream [7] divides given streaming data into multiple sub-streams and delivers those sub-streams using a forest of trees, one for each sub-stream, trying to use each peer as an interior node in at most one tree and as a leaf node in remaining trees. This is developed with the aid of Scribe [16] which is known as an application-level group communication scheme based on a DHT-based P2P overlay called Pastry [17]. Each group, in Scribe, is given a pseudo-random Pastry key as a group Id, and trees are built using reverse path forwarding on the union of Pastry routes from each group member to the roots. More precisely, in each step, a peer forwards a message to a peer whose Id shares a longer prefix with the group Id, i.e., the root. Thus, by choosing group Ids for the trees that all differ in the most significant digit, SplitStream ensures that trees have a disjoint set of interior peers.

When a peer joins the SplitStream system, it selects random peers, i.e., random Pastry Ids, and asks them to be their parents. If a peer can not adopt more children and receives a join request from another peer, one of its children will be rejected according to the value of a utility function. Rejected child seeks for another parent by referring to a set of peers with excess capacity called the spare capacity group. Thus, SplitStream is able with a high probability to reorganize a collection of trees even when the upload capacity of each peer is fully used, and hence broadcasting the maximum streaming rate for a large population. However, especially in our case of research when trying to maximally utilize the upload capacity of each peer, i.e., the spare capacity is low, the mechanism of rejecting children will frequently happen, and in many cases, peers will not find a peer as a parent before asking the spare capacity group; This will be verified in our simulations. Thus, the search for spare capacity peers happens frequently which is a time consuming process. The more important consequence of asking the spare capacity peers is that those peers will have a small number of children in different trees leading to degenerate trees.

## 3. PROPOSED SCHEME

### 3.1. PRELIMINARIES

Notions used in the proposed scheme are summarized in Table 1. In the proposed scheme, we divide the given video stream with bit rate $r$ into $N$ sub-streams of bit rate $s = r/N$ each, and deliver those sub-streams through different spanning trees. Thus in the following, we will use terms "tree" and "sub-stream" interchangeably.

Let $V$ be a set of $n$ peers and $\mathcal{T} = \{T_1, T_2, \ldots, T_N\}$ be a variable set of $N$ trees (sub-streams). Each peer $i \in V$ can have different number of children in each tree in $\mathcal{T}$, while the total number of children should not exceed a value $m(i)$ determined by the upload capacity of the peer. In the following, we call $m(i)$ the ***budget*** of peer $i$, and will design a scheme such that the role of uploading a sub-stream is transferred to another peer by exchanging money among peers, provided that the balance of each peer is not below zero.

Given a collection of trees $\mathcal{T}$, the ***price*** of peer $i$ with respect to the $k^{th}$ tree $T_k$ is defined as the number of children of $i$ in $T_k$ plus one. Such $N$ prices of peer $i$ are locally stored in the form of a price vector $C_i$ of length $N$. Note that $C_i[k] \geq 1$ for any $i$ and $k$. A peer is said to be ***saturated*** if it has the maximum number of children in only one tree. More particularly, peer $i$ is saturated with respect to the $k^{th}$ tree $T_k$ if $C_i[k] = m(i) + 1$ and $C_i[h] = 1$ for all $h \neq k$. The tree $T_k$ is said to be a dominant sub-stream for peer $i$ if $C_i[k] = max_k C_i[k]$.

### 3.2. BASIC OPERATIONS

Suppose that each peer is associated with a set of $D$ random peers (neighbours) by the tracker. Let $U_i$ be a subset of peers associated with the peer $i$. In the proposed scheme, peer $i$ can subscribe to a (new) sub-stream by communicating with peers $j$ in $U_i$. The concrete scheduling algorithm, the detail of which will be described in Section 3.3, is based on three ways of reconfiguring trees in $\mathcal{T}$ (Figure 1). The three ways are designed to increase the number of saturated peers. If $i$ could not finish the scheduling due to the lack of resources in $U_i$, it contacts peers in a set of peers with free capacity, the detail of which is described in Section 3.5.

**Way-1**: The first way of reconfiguring $\mathcal{T}$ is to use the free upload capacity of a peer. More concretely, if peer $j \in U_i$ is subscribing to the $k^{th}$ sub-stream and has a free upload capacity, then peer $i$ can subscribe to the $k^{th}$ sub-stream by making itself as a child of $j$ in $T_k$ (note that such an action decreases the balance of $j$ by one). See Figure 1(A) for illustration. If there are several such pairs of parent and sub-stream, $i$ prefers to a pair of $j$ and $k$ such that $k$ is a dominant sub-stream of $j$. That is because the join of $i$ to $j$ in $T_k$ increases the value of $C_j[k]$ by one which makes $j$ closer to saturation.

Table 1. Main notions in this paper.

| | |
|---|---|
| $n$: | number of peers |
| $N$: | number of sub-streams (trees) |
| $r$: | streaming rate |
| $s$: | sub-stream rate |
| $m(i)$: | budget of peer $i$ (maximum number of children) |
| $C_i$: | price vector of peer $i$ |
| $C_i[k]$: | price of peer $i$ in $k^{th}$ tree |
| $\beta(i)$: | balance of peer $i$ |
| $U_i$: | set of neighbours of peer $i$ |
| $D$: | number of neighbours |
| $P_i$: | set of sub-streams not subscribed by peer $i$ |

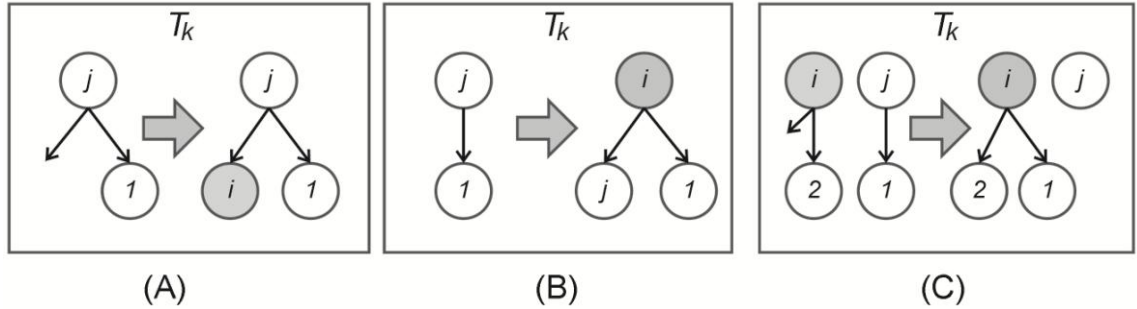

Figure 1. Three ways for reconfiguring a tree $T_k$. (A) peer $i$ can be a child of $j$ simply because $j$ has a free capacity. (B) peer $i$ can buy a sub-stream $T_k$ from peer $j$ by paying a price of 2, and hence peer $i$ has two children. (C) both peers $i$ and $j$ have one child in the same tree $T_k$ so peer $i$ asks $j$ to swap his child.

**Way-2**: The second way is to buy the right of uploading sub-streams by paying money. More particularly, if $j$ is subscribing to the $k^{th}$ sub-stream and $i$ has a positive balance, then by paying money of amount $m(\geq 1)$ to $j$, $i$ can subscribe to the $k^{th}$ sub-stream by taking the place of $j$ in $T_k$ and is granted the right of uploading the sub-stream to $j$ and its $m - 1$ children. Figure 1(B) illustrates this case. The only restriction in this case is that $j$ must not be saturated with respect to the transferred sub-stream as that reduces the number of saturated peers.

**Way-3**: The third way is to swap children with other peers. Suppose that there are two peers $i$ and $j$ subscribing to the $k^{th}$ sub-stream, where: 1) $i$ has a free upload capacity while $T_k$ is its dominant sub-stream and 2) $j$ has at least one child in $T_k$ but it is not a dominant sub-stream for $j$. Then, $i$ asks $j$ to hand over the right of uploading the $k^{th}$ sub-stream to one child of $j$ in $T_k$ by paying one unit of money. This way is exemplified in Figure 1(C).

### 3.3. SCHEDULING PROCESS

In the proposed algorithm, peers subscribe to sub-streams through two phases. The role of the first phase is to increase the number of peers that are internal in only one tree, i.e., to increase the number of saturated peers. The second phase is just to complete the subscription to all sub-streams.

**First Phase**: Let $P_i$ be a variable representing the set of sub-streams which is not subscribed by the peer $i$. For each $q \in P_i$, peer $i$ counts the neighbours who have children in tree $q$ but $q$ is not their dominant sub-stream. Then, those neighbors are willing to reject children in $q$ and adopt children in their dominant sub-stream. Let $q^* \in P_i$ be a sub-stream with a maximum count and $U_i^* (\subseteq U_i)$ be the set of neighbours contributing to the count of $q^*$. To increase the number of peers uploading exactly one sub-stream, peer $i$ conducts the following two steps sequentially:

**Step 1:** Let $\beta(i)$ be the balance of peer $i$. This step is executed only when $\beta(i) \geq 1$. Peer $i$ selects a peer $j \in U_i^*$ with a shortest depth from the root in the tree corresponding to $q^*$, and buys $q^*$ from $j$ by paying money. Peer $i$ either pays one unit of money and subscribes to $q^*$ (Way-2) or pays more than one unit of money and subscribes to the dominant sub-stream of $j$ (Way-1) in addition to $q^*$ (Way-2). In the latter case, the paid money should not exceed $\beta(i) - (|P_i| - 2)$, since we need to reserve money for the remaining $|P_i| - 2$ unsubscribed sub-streams. The reader should note that $q^*$ is the potential dominant sub-stream for peer $i$.

**Step 2:** For each $j' \in U_i^* \setminus \{j\}$, peer $i$ gets one child of $j'$ for sub-stream $q^*$ (Way-3) and subscribes to the dominant sub-stream of $j'$ (Way-1), if $i$ has not yet subscribed to it. Note that sub-stream $q^*$ should be commonly subscribed by $i$ and $j'$ but is not a dominant sub-stream of $j'$. The last option for peer $i$ to increase the number of children for the dominant sub-streams of peers in $U_i$ is to look for peers that have a free capacity and subscribe to their dominant sub-streams. Thus, we have the third step:

**Step 3:** If there is a peer $j \in U_i$ such that $j$ has a free capacity and $i$ has not subscribed to a "dominant" sub-stream $q$ of $j$, then $i$ becomes a child of $j$ with respect to $q$. This operation is repeated until there is no such peer $j$ in $U_i$.

**Second Phase**: The second phase is executed when peer $i$ could not subscribe to all sub-streams after the first phase being finished. At any step in this phase, if $P_i$ becomes empty, peer $i$ proceeds to Step 6 to increase the number of children for its dominant sub-stream. The steps are as follows.

**Step 4:** For each unsubscribed sub-stream $q \in P_i$, peer $i$ seeks a peer $j \in U_i$ such that $C_j[q]$ is the cheapest among all peers in $U_i$ and $q$ is not the dominant sub-stream of $j$. Then, peer $i$ buys $q$ from $j$ by paying money (Way-2). A draw in prices is resolved by the hop-count delay. Peer $i$ looks for the cheapest price to save the money to get more children in its dominant sub-stream $q^*$.

**Step 5:** At this point, the budget of peer $i$ is exhausted. Thus in order to subscribe to a new sub-stream in $P_i$, $i$ needs to use the free capacity of other peers in $U_i$ (Way-1). Recall that the use of the free capacity of peer $j$ does not decrease the balance of $i$, but it decreases the balance of $j$ because it reduces the amount of free capacity of $j$. If there is no peer with an available free capacity in $U_i$, as a last resort, peer $i$ asks peers in the free set until $P_i$ becomes empty (the way of maintaining the free set is described in subsection 3.5).

**Step 6** (Post Processing): If $m(i) > 0$ and $U_i^* \neq \emptyset$ at this point, peer $i$ tries to collect as many children for sub-stream $q^*$ as possible from peers in $U_i^*$. We need to notice that this is a special case of the swap process (Way-3) so that no new subscription occurs, and is conducted only once.

### 3.4. INITIALIZATION

In the following, we describe the initialization of the system by the proposed algorithm using a simple example. In the example, all peers have a uniform upload capacity of four, i.e., $u_j = 4$ for each $j$, and the given video stream is divided into four sub-streams of unit bit rate ($s = 1$), i.e., $r = 4$ and $N = 4$. Peer $i$, with an upload capacity $u_i = 4$, wants to join the system. The budget of peer $i$ is determined as $m(i) := u_i/s = 4$ and the set of neighbours is given as $U_i = \{a, b, c, d\}$. Given a collection of trees $\mathcal{T}$ shown in Figure 2(A), price vectors are calculated as shown on top of the figure. From such vectors, we notice that: 1) peers $b$ and $c$ are saturated in the fourth and the third trees, respectively, and 2) peers $a$ and $d$ have the second sub-stream as a dominant one.

From the Figure 2(A), we notice that the first sub-stream is not dominant for both peers $a$ and $d$ with a price equals to two. Thus, peer $i$ selects the first sub-stream as $q^*$ along with peers $a$ and $d$ be the members of $U_i^*$. According to step 1, Figure 2(B), peer $i$ buys $q^*$ from the peer $a$ ($\in U_i^*$) by paying two units of money. That means peer $i$ will replace peer $a$ in the first tree and adopt both peer $a$ and his child. Then, peer $a$ has got a free capacity by receiving money from peer $i$. That allows peer $a$ to adopt peer $i$ in the second tree corresponding to the dominant sub-stream of peer $a$.

In Figure 2(C), representing step 2, peer $i$ asked peer $d$ to swap its child in first tree. However, it could not subscribe to the dominant sub-stream of peer $d$, which is the second tree, as it is already subscribed to. Note that by this action the balance of peer $i$ is reduced by one and peer $d$ has got a free capacity. At this point, as peer $d$ is the only peer that has a free capacity and its dominant sub-stream is not required by peer $i$, the step 3 will have no effect on the overlay.

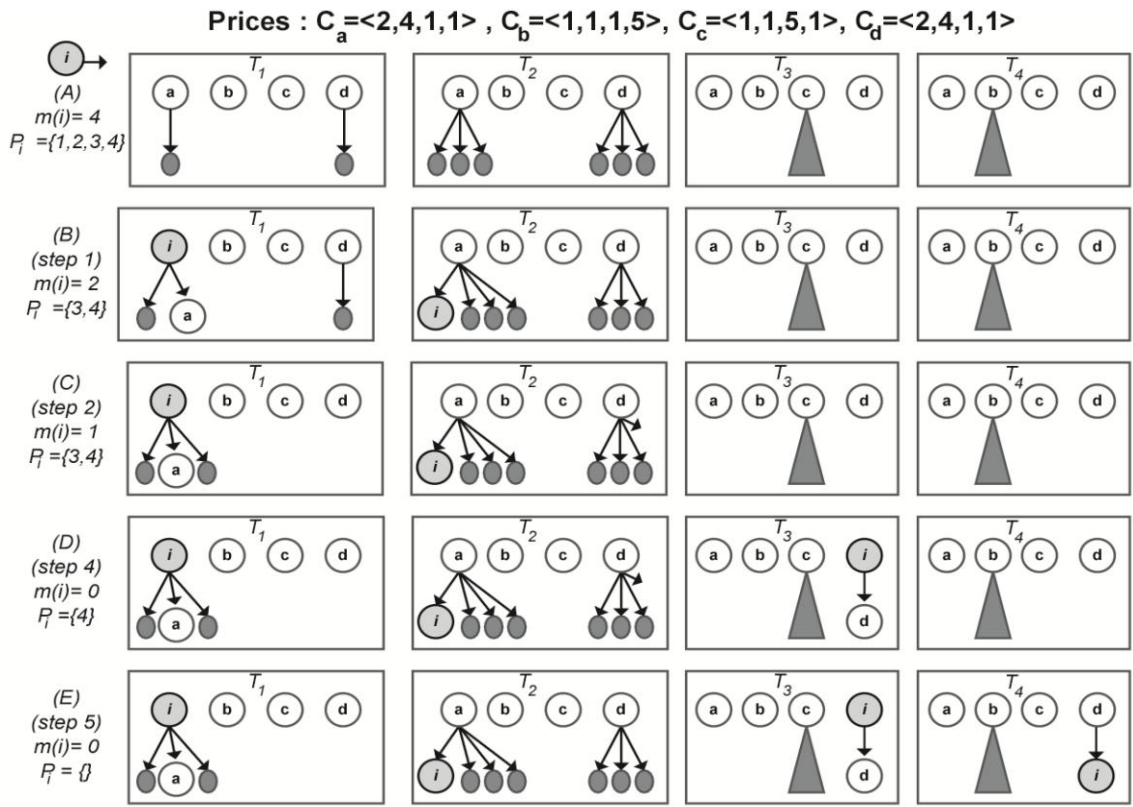

Figure 2. A scheduling example. The triangle means that the peer has all its children in this tree.

In the second phase of the algorithm, peer $i$ starts with step 4, illustrated in Figure 2(D). There are three peers to have a price equals to one in the third tree, and peer $i$ chooses peer $d$ as the seller of the third sub-stream (note that hop-counts are not illustrated in this figure for simplicity). By choosing peer $d$, peer $i$ will replace it in the third tree and adopt it by paying one unit of money. Finally, Figure 2(E) illustrates the case of step 5 in which peer $i$ becomes a child of peer $d \in U_i$ in the fourth tree. The post processing step will be skipped by peer $i$ as its balance is zero.

### 3.5. HOW TO MANAGE THE FREE SET

To implement the free set in a distributed environment, we get benefit from the forest of trees organized in the proposed scheme. In this subsection, we describe a concrete way to realize three operations used for the free set, i.e., join, leave and find.

To join the free set, a peer $i$ tells all its parents about that (recall that it has at most $N$ parents in $\mathcal{T}$). After receiving a message from a child in a tree, each peer forwards the information to the parent in the tree unless it is the root. As a result, we have a path from $i$ to the root in each tree so that all peers on the path are aware of that $i$ is a member of the free set. This operation takes at most $N \times d$ messages provided that the maximum depth of trees is bounded by $d$. The leave from the free set is conducted in a similar way. If peer $i$ in the free set changes the parent in a tree, which frequently occurs in the scheduling process, such an update must be propagated to all peers on the paths by the old and new parents of $i$, i.e., the old parent initiates the propagation of leave message and the new parent initiates the propagation of join message.

If peer $j$ wants to find a peer in the free set, it sends a request message to one of its parents selected randomly. The request is forwarded up in the corresponding tree until it finds a peer that knows about one of the free set peers. Note that such a forwarding process can always find a peer in the free set in at most $d$ hops (if any), since the root of any tree knows all members of the free set. The reader should note that in the above process, the root of a tree does not become a bottleneck in many cases, because: 1) the tree is randomly selected from $N$ candidates in $\mathcal{T}$ and 2) it is likely that a request path and a join path will meet at a deep level of the selected tree. If a peer in the free set receives several requests from different peers, it serves its upload capacity in the first-come and first-serve basis.

## 4. EVALUATION

To evaluate the performance of the proposed scheme, we conducted extensive simulations based on OPSS [18]. The performance of the scheme is compared with SplitStream where not to reinvent the wheel, we used an OPSS simulation package developed for SplitStream [19] in the evaluation. The efficiency of constructed multiple-tree overlays is evaluated through the following three metrics:

- *Saturation fraction* of a peer is the ratio of the number of children for its dominant substream to the maximum number of children of the peer (i.e., budget). It takes a value in range [0, 1] where a higher value implies that the leave of the peer affects its descendants for a smaller number of sub-streams.
- The hop-count delay of a peer in a tree is the number of links on the unique path connecting the peer to the root (source) in the tree. The *average hop-count* of a peer indicates the average of the hop-count delay over all trees.
- *Free set requests* represent the total number of requests received by the free set to complete a scheduling. We are interested in this metric due to the fact that the maintenance cost of the free set and the cost required for seeking subscribers heavily affect the overhead of the scheme. More importantly, by asking the peers of the free set, those peers will have a small number of children in different trees leading to degenerate trees.

## 4.1. SETUP

In the following, we represent the upload capacity of peers in terms of the budget and the rate of video streams in terms of the number of sub-streams, i.e., we normalize actual values by the bit rate of sub-streams. Table 2 summarizes all scenarios examined in the evaluations, where in each scenario, the download capacity of each peer is assumed to be sufficiently large. Scenario's name in the table is encoded by the environment type, HM (homogeneous) or HT (heterogeneous), followed by the bit rate of given video stream (e.g., 4 means that the stream is divided into four sub-streams), and the resource index where 1 stands for $R = 1.0$ and 2 stands for $R = 1.25$. The reader should note that the resource index, $R$, is defined as the ratio of the available capacity in the system to the streaming rate times the number of peers as in [7].

Heterogeneous settings follow the setting used in [20]. More concretely, we adopt three types of upload capacities low, medium and high which correspond to the bit rate of 128 [Kbps], 384 [Kbps] and 1000 [Kbps], respectively, and we fix the sub-stream rate to either 64 [Kbps] or 128 [Kbps]; thus in the former case, the upload capacity of each type is normalized to 2, 6 and 16, respectively. The fraction of each type of peers in the population is fixed as in Table 3.

Table 2. Simulation scenarios.

|  | HM4-1 | HM4-2 | HM8-1 | HM8-2 | HT4-1 | HT4-2 | HT8-1 | HT8-2 |
|---|---|---|---|---|---|---|---|---|
| **Server capacity** | 4 | 5 | 8 | 10 | 4 | 4 | 8 | 8 |
| **Peer capacity** | 4 | 5 | 8 | 10 | 1,3,8 | 1,4,10 | 2,6,16 | 3,7,20 |
| **Stream rate** | 4 | 4 | 8 | 8 | 4 | 4 | 8 | 8 |
| **Resource Index** | 1 | 1.25 | 1 | 1.25 | 1 | 1.25 | 1 | 1.25 |

For each scenario, we ran the proposed scheme and SplitStream by fixing the number of peers to $n = 10000$, where we did not consider churn to make a fair comparison of the schemes. The proposed scheme is evaluated for different values of $D$ (the number of peers in set $U_i$). Although $D$ was chosen to be a multiple of the number of sub-streams (namely, $N$, $2N$ or $4N$) in the simulation, any value can be used for $D$. The saturation fraction and the average hop-count delay are calculated for each peer and the cumulative distributions are plotted for only some scenarios to save the space. On the other hand, the average value over all peers is presented in tables for all scenarios.

Table 3. Fraction of each type in the population.

| Type | Fraction |
|---|---|
| Low (128 Kbps) | 37% |
| Medium (384 Kbps) | 27% |
| High (1000 Kbps) | 36% |

## 4.2. RESULTS

### 4.2.1. SATURATION FRACTION

As was mentioned, a peer with a high saturation fraction means a lower number of sub-streams to be lost in case of its leave. However, as will be seen in the next section, a higher saturation fraction does not necessarily mean a shorter hop-count delay, since a nearly-saturated peer might have few children as leaves in other trees.

Figure 3 shows the cumulative distribution of the saturation fraction with different number of neighbors, $D$, and Table 4 summarizes the average saturation fraction in each scenario, where (SS) stands for SplitStream.

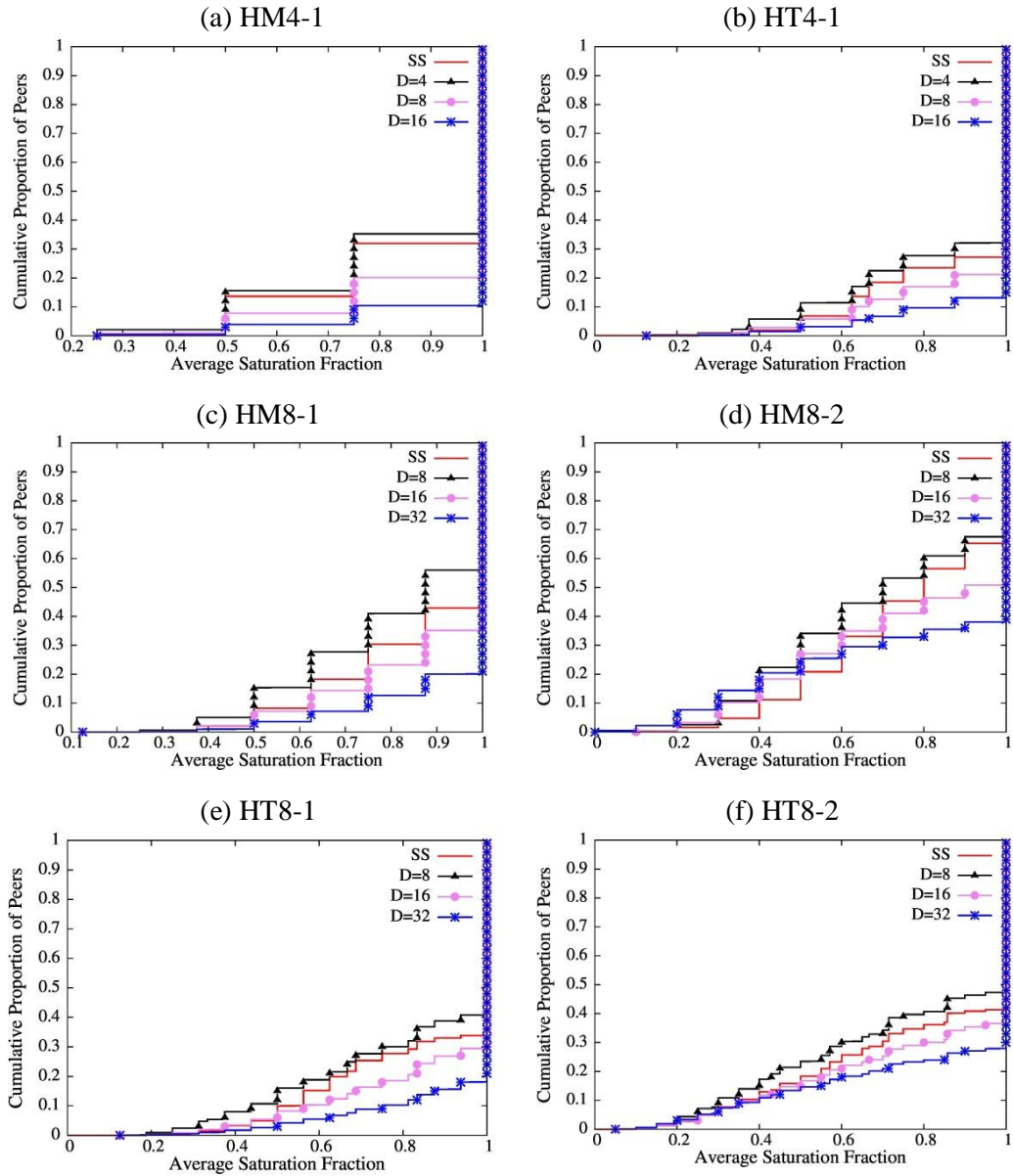

Figure 3. Cumulative distribution of saturation fraction

The average saturation fraction of the proposed scheme increases as $D$ increases since it increases the chance of buying or swapping sub-streams with other peers, whereas it is worse than Split-Stream when $D$ takes the smallest value $N$. From Figure 3, we can also confirm that the number of peers that have their children in more than one tree is 10% in HM4-1 and 20% in HM8-1 for $D = 4N$. It means that the saturation fraction is higher for a lower number of sub-streams. It should also be noted that the saturation fraction degrades by increasing the resource index $R$ from 1.0 to 1.25. In fact, if $R$ is sufficiently large, it is possible to attain the given streaming rate without fully utilizing upload capacities, which prevents many peers from being saturated.

Table 4. Average saturation fraction.

|        | SS   | D = 4 | D = 8  | D = 16 |
|--------|------|-------|--------|--------|
| HM4-1  | 0.89 | 0.87  | 0.93   | 0.96   |
| HM4-2  | 0.76 | 0.73  | 0.78   | 0.80   |
| HT4-1  | 0.90 | 0.88  | 0.93   | 0.96   |
| HT4-2  | 0.82 | 0.80  | 0.85   | 0.87   |
|        | SS   | D = 8 | D = 16 | D = 32 |
| HM8-1  | 0.87 | 0.82  | 0.90   | 0.94   |
| HM8-2  | 0.76 | 0.70  | 0.77   | 0.79   |
| HT8-1  | 0.87 | 0.84  | 0.90   | 0.94   |
| HT8-2  | 0.81 | 0.78  | 0.84   | 0.86   |

**4.2.2. AVERAGE HOP-COUNT**

Next, we evaluate the average hop-count delay of the proposed scheme. Figure 4 shows the cumulative distribution of the average hop-count delay and Table 5 shows its average in each scenario, as before. We can observe that the proposed scheme outperforms SplitStream for all scenarios, and in seven out of eight scenarios, it attains a shorter hop-count delay than SplitStream by at least 1.0 even when $D = N$ (note that the difference becomes large for large $D$'s). Figure 4 also clarifies that the proposed scheme outperforms SplitStream with respect to the "maximum" average hop-count delay.

Such a positive effect of parameter $D$ reduces for large resource index $R$. In fact, in scenario HM4-2 with $R = 1.25$, the average hop-count increases as $D$ increases in contrast to other scenarios with $R = 1.25$. One possible conjecture to explain such a phenomenon is that for large $R$'s, as $D$ increases, the average hop-count decreases up to a limit related to the number of sub-streams and after that limit, the delay increases again due to the (unnecessary) join to other trees as leaves of deeper level.

To verify this conjecture, we conducted additional simulation for HM8-2 and HM8-1 and increased $D$ up to 48. As a result, we found that the average hop-count of HM8-2 increases from 4.99 to 5.31 by increasing $D$ from 40 to 48, and that of HM8-1 does not change from 5.47 regardless of the increase of $D$ (note that it has almost reached the optimal value since a theoretical bound for HM8-1 is 5.465).

Table 5. Average hop-count

|        | SS    | D = 4 | D = 8  | D = 16 |
|--------|-------|-------|--------|--------|
| HM4-1  | 9.58  | 7.43  | 7.31   | 7.29   |
| HM4-2  | 7.96  | 6.62  | 6.47   | 7.03   |
| HT4-1  | 12.51 | 8.52  | 8.43   | 8.22   |
| HT4-2  | 9.95  | 8.05  | 7.60   | 7.45   |
|        | SS    | D = 8 | D = 16 | D = 32 |
| HM8-1  | 6.53  | 5.52  | 5.43   | 5.47   |
| HM8-2  | 5.54  | 5.05  | 4.95   | 4.93   |
| HT8-1  | 7.52  | 5.60  | 5.47   | 5.37   |
| HT8-2  | 6.28  | 5.28  | 5.07   | 5.05   |

Another important issue to address is that why the proposed scheme outperforms SplitStream even under a low saturation fraction? To clarify this point, we analyzed the difference of the structure of the resulting multiple-trees to an optimal multiple-tree, which can be obtained for homogeneous cases as follows.

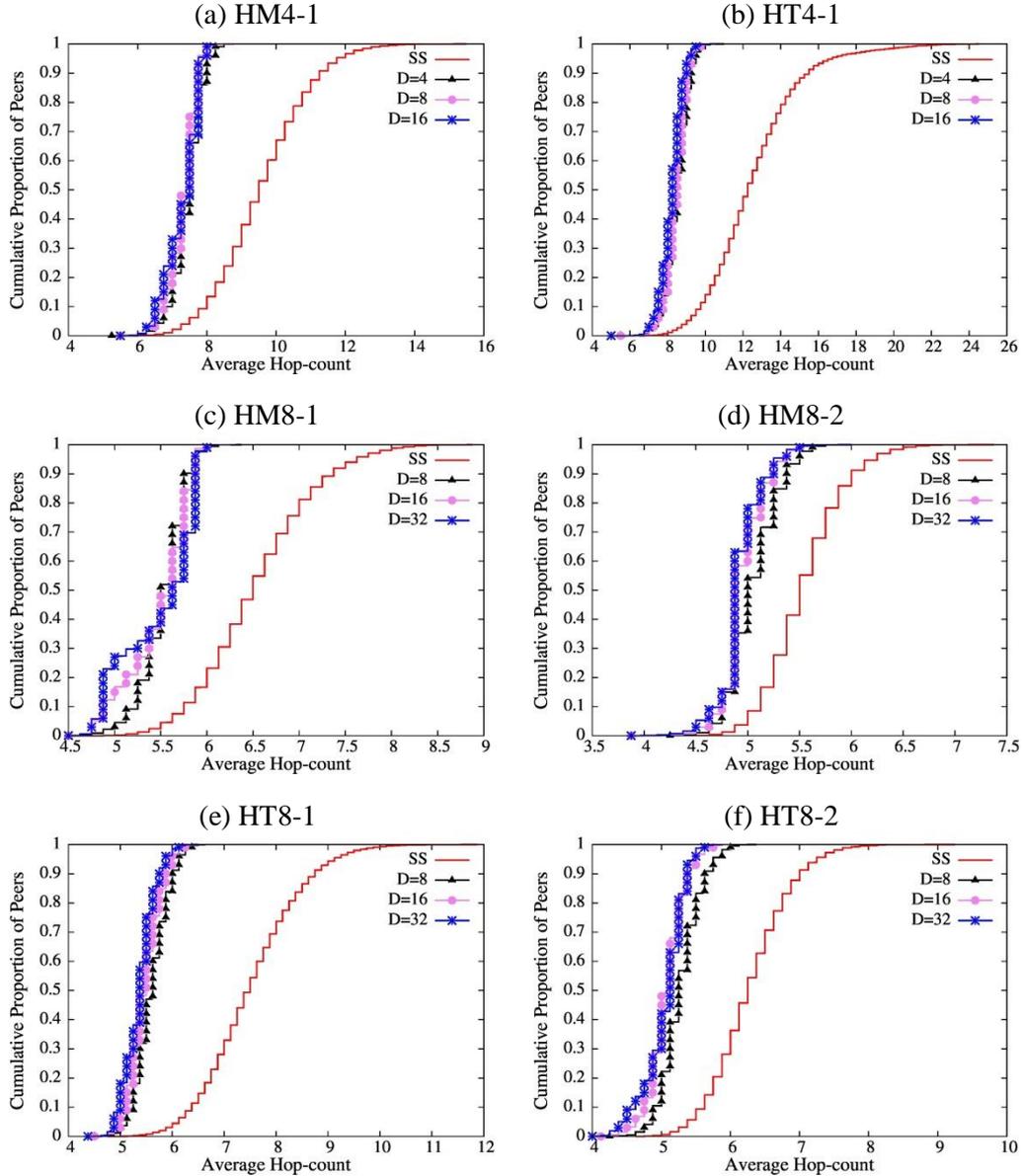

Figure 4. Cumulative distribution of average hop-count.

Since the number of children of each peer is bounded by $N$, an optimal tree contains $N^{l-1}$ peers at the $l^{th}$ level for each $l$ (e.g., the first level consists of the root of the tree, the second level consists of $N$ children of the root, and so on) except for the deepest level of the tree, where the depth $d$ of the tree can be obtained by solving $\sum_{i=1}^{d-1} N^{i-1} < n \leq \sum_{i=1}^{d} N^{i-1}$ which is approximately $log\{n(N-1)+1\}/log\ N$.

The number of uploaders (peers with at least one child) at the $l^{th}$ level of the optimal tree can thus be calculated as follows:

1) for $1 \leq l < d - 1$, it is $N^l$ and,
2) for $l = d - 1$, it is $\left\lceil \frac{n - \sum_{i=1}^{d-1} N^{i-1}}{N} \right\rceil$.

Consequently, the number of uploaders at the $l^{th}$ level across all trees in an optimal multiple-tree is given as

$$N \times min\left\{N^{l-1}, \left\lceil \frac{n - \sum_{i=1}^{l} N^{i-1}}{N} \right\rceil\right\} \qquad (2)$$

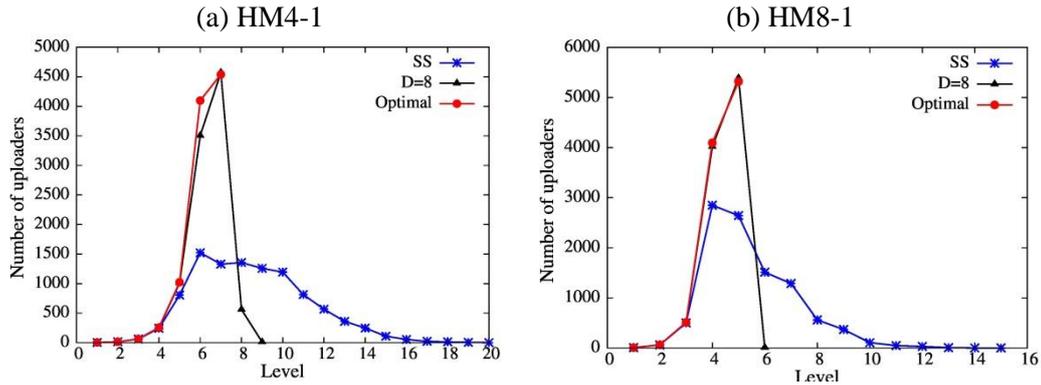

Figure 5. Number of uploaders in different levels.

Figure 5 compares the resulting multiple-trees with an optimal one, for scenarios HM4-1 and HM8-1. The horizontal axis of the figure is the level of the tree and the vertical axis is the number of uploaders at each level. The proposed scheme matches the optimal tree up to the fifth level, and it is nearly optimal even for deeper levels. On the other hand, SplitStream goes far from optimal with a large gap (e.g., the gap which is 2000 uploaders at the sixth level in HM4-1) and with the existence of many uploaders at deeper levels. Recall that the proposed scheme has been designed to increase the number of saturated peers in all trees. Moreover, as peers prefer to buy or swap other peers that have a short hop-count delay in case of a price draw, the proposed scheme can maintain short depth trees. As for SplitStream, the random selection of parents according to Pastry Id can not guarantee an efficient overlay construction with a short hop-count delay and leads to a high use of the peers in the free set capacity, as will be verified in the next subsection, resulting in this kind of degenerate trees.

### 4.2.3. FREE SET REQUESTS

Finally, we evaluate the amount of free set requests issued by the participants. Figure 6 shows the fraction of peers which issued (at least one) free set request before completing the scheduling. Recall that such a request is issued when it does not have enough balance or it can not find a neighbor which has enough upload capacity. In homogeneous scenarios, the proposed scheme causes no free set request, whereas the fraction of peers which issue a free set request in SplitStream is 60% for $R = 1.0$ and 30% for $R = 1.25$. This means that, in homogeneous environment, the proposed scheme is remarkably efficient compared with SplitStream with respect to the overhead for the maintenance of free set.

The superiority of the proposed scheme to SplitStream can be observed even under heterogeneous scenarios provided that $R = 1.25$ and such an effect is enhanced for larger $D$'s. For example, in HT8-2, exactly one peer (among 10000 peers) issued a free set request for $D = 32$.

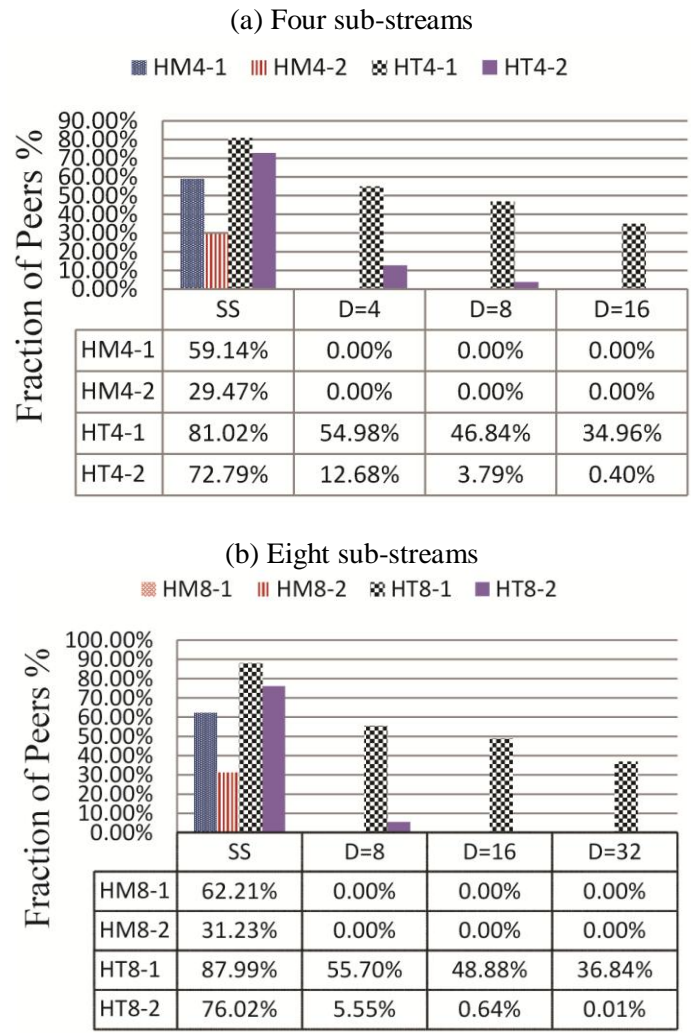

Figure 6. Free set requests

## 5. CONCLUDING REMARKS

This paper proposes a scheduling scheme for P2P streaming systems which attains the maximal resource utilization in a distributive manner. The proposed scheme is able to build an efficient multiple-tree overlay with a short hop-count delay even when no spare capacity is available. The result of simulation proves that: 1) the constructed multiple-trees certainly converge to anefficient overlay with a short hop-count delay, and 2) it outperforms SplitStream with respect to the average hop-count in all scenarios examined in the experiments. In addition, 3) the proposed scheme outperforms SplitStream in regard to the number of peers who are internal in only one tree (saturated peers) provided that the number of allowed neighbors is more than the number of sub-streams.

## REFERENCES

[1] Cisco Systems Inc. (2015) Cisco Visual Networking Index: Forecast and Methodology, 2014-2019, http://www.cisco.com/c/en/us/solutions/collateral/service-provider/ip-ngn-ip-next-generation-network/white_paper_c11-481360.pdf May. 2015.

[2] X. Zhang, J. Liu, B. Li, and P. Yum, (2005) "CoolStreaming/DONet: A Data-driven Overlay Network for Peer-to-Peer Live Media Streaming," Proceedings of IEEE Infocom, pp.2102-2111.

[3] M. Zhang, Q. Zhang, L. Sun, and S. Yang, (2007) "Understanding the Power of Pull-Based Streaming Protocol: Can We Do Better?," IEEE J.Sel. A. Commun., vol.25, no.9, pp.1678-1694.

[4] Y. Chu, S.G. Rao, and H. Zhang, (2000) "A case for end system multicast," Proceedings of. ACM Sigmetrics '00, pp.1-12.

[5] S. Zhuang, B. Zhao, A. Joseph, R. Katz, and J. Kubiatowicz, (2001) "Bayeux: an architecture for scalable and fault-tolerant wide-area data dissemination," Proceedings of NOSSDAV'1, pp.11-20,.

[6] S. Ratnasamy, M. Handley, R. Karp, and S. Shenker, (2001) "Application-Level Multicast Using Content-Addressable Networks," Proceedings of NGC'1, pp.14-29.

[7] M. Castro, P. Druschel, A.-M. Kermarrec, A. Nandi, A. Rowstron, and A. Singh, (2003) "SplitStream: High-bandwidth content distribution in cooperative environments," Proceedings of SOSP'03, pp.298-313.

[8] V. Padmanabhan, H. Wang, P. Chou, and K. Sripanidkulchai, (2002) "Distributing Streaming Media Content using Cooperative Networking," Proceedings of NOSSDAV'2, pp.177-186.

[9] R. Sweha, V. Ishakian, and A. Bestavros, (2012) "AngelCast: Cloud-based Peer- Assisted Live Streaming Using Optimized Multi-Tree Construction, " Proceedings of ACM MMSys, pp.191-202.

[10] H. Yin, X. Liu, T. Zhan, V. Sekar, F. Qiu, C. Lin, H. Zhang, and B. Li, , (2009) "Design and Deployment of a Hybrid CDN-P2P System for Live Video Streaming: Experience with LiveSky," Proceedings of 17th ACM international conference on Multimedia , pp.25-34.

[11] R. Kumar, Y. Liu, and K.W. Ross, (2007) "Stochastic Fluid Theory for P2P Streaming Systems," Proceedings of IEEE Infocom, pp.919-927.

[12] Y. Guo, C. Liang, and Y. Liu, (2008) "AQCS: Adaptive Queue-Based Chunk Scheduling for P2P Live Streaming," Proceedings of IFIP Networking, pp.433-444.

[13] S. Sengupta, S. Liu, M. Chen, M. Chiang, J. Li, and P. A. Chou, (2011) "Peer-to-Peer Streaming Capacity," IEEE Trans. Inf. Theory, vol. 57, no. 8, pages 5072-508.

[14] S. Zhang, Z. Shao, M. Chen, and L. Jiang, (2014) "Optimal Distributed P2P Streaming under Node Degree Bounds," IEEE/ACM Trans. Networking, vol. 22, no. 3, pages 717-730.

[15] M. S. Raheel, R. Raad, and C. Ritz, (2015) "Achieving maximum utilization of peer's upload capacity in p2p networks using SVC," Peer-to-Peer Netw. Appl.

[16] M. Castro, P. Druschel, A.-M Kermarrec, and A. Rowstron, (2006) "SCRIBE: A large-scale and decentralized application-level multicast infrastructure," IEEE J.Sel. A. Commun., vol.20, no.8, pp.1489-1499.

[17] A. Rowstron and P. Druschel, (2001) "Pastry: Scalable, distributed object location and routing for large-scale peer-to-peer systems," Proceedings of IFIP/ACM Middleware '01, pp.329-350.

[18] L. Bracciale, F. Lo Piccolo, D. Luzzi, and S. Salsano, (2007) "OPSS: an overlay peer-to-peer streaming simulator for large-scale networks," SIGMETRICS Perform. Eval. Rev., vol.35, no.3, pp.25-27.

[19] G. Bianchi, N. B. Melazzi, L. Bracciale, F. Lo Piccolo, and S. Salsano, (2010) "Streamline: An Optimal Distribution Algorithm for Peer-to-Peer Real-Time Streaming," IEEE Trans. Parallel Distrib. Syst., vol.21, no.6, pp.857-871.

[20] S. Saroiu, K. P. Gummadi, and S. D. Gribble, (2002) "A Measurement Study of Peer-to-Peer File Sharing Systems," Proceedings of Multimedia Computing and Networking (MMCN).

**Authors**

**Bahaa Aldeen ALGHAZAWY** received the B.E. degree in electronic engineering from University of Aleppo in 2009, and the M.E. degree in information engineering form Hiroshima University in 2013. He is currently a Ph.D. candidate at the Department of Information Engineering, Hiroshima University. His research interests are in the area of internet and peer-to-peer networks with emphasis on media streaming.

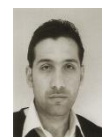

**Satoshi FUJITA** received the B.E. degree in electrical engineering, M.E. degree in systems engineering, and Dr.E. degree in information engineering from Hiroshima University in 1985, 1987, and 1990, respectively. He is a Professor at Graduate School of Engineering, Hiroshima University. His research interests include communication algorithms, parallel algorithms, graph algorithms, and parallel computer systems. He is a member of the Information Processing Society of Japan, the Institute of Electronics, Information and Communication Engineers, SIAM Japan, IEEE Computer Society, and SIAM. 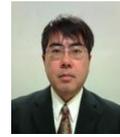